\documentclass{article}

 \usepackage[preprint]{neurips_2025}

\usepackage[utf8]{inputenc} 
\usepackage[T1]{fontenc}    
\usepackage{hyperref}       
\usepackage{url}            
\usepackage{booktabs}       
\usepackage{amsfonts}       
\usepackage{nicefrac}       
\usepackage{microtype}      
\usepackage{xcolor}         
\usepackage{graphicx}
\usepackage{setspace}
\PassOptionsToPackage{numbers, compress}{natbib}
\title{Lattica: A Decentralized Cross-NAT Communication Framework for Scalable AI Inference and Training}

\author{
  Ween Yang\textsuperscript{\rm 1}\thanks{Equal contribution}, 
    Jason Liu\textsuperscript{\rm 1}\footnotemark[1], 
    Suli Wang\textsuperscript{\rm 1,2}\footnotemark[1],
    Xinyuan Song\textsuperscript{\rm 1,3}\footnotemark[1],\\
    Lynn Ai\textsuperscript{\rm 1},
    Eric Yang\textsuperscript{\rm 1},
    Bill Shi\textsuperscript{\rm 1}\thanks{Corresponding author} \\
  \textsuperscript{\rm 1}Gradient\\
  \textsuperscript{\rm 2}Technische Universität Darmstadt, Germany\\
  \textsuperscript{\rm 3}Emory University, USA\\  
  \texttt{tianyu@gradient.network} \\
}

\begin{document}

\maketitle

\begin{abstract}
\renewcommand{\thefootnote}{\fnsymbol{footnote}}
The rapid expansion of distributed Artificial Intelligence (AI) workloads beyond centralized data centers creates a demand for new communication substrates. These substrates must operate reliably in heterogeneous and permissionless environments, where Network Address Translators (NATs) and firewalls impose significant constraints. Existing solutions, however, are either designed for controlled data center deployments or implemented as monolithic systems that tightly couple machine learning logic with networking code. To address these limitations, we present \textbf{Lattica}\footnote{Gradient Network, \textit{Introducing Lattica: The Universal Data Motion Engine} [EB/OL]. 2025-06-19 [2025-09-30]. Available at: \url{https://gradient.network/blog/lattica-universal-data-motion-engine}.}, a decentralized cross-NAT communication framework designed to support distributed AI systems. Lattica integrates three core components.  
First, it employs a robust suite of NAT traversal mechanisms to establish a globally addressable peer-to-peer mesh. Second, it provides a decentralized data store based on Conflict-free Replicated Data Types (CRDTs), ensuring verifiable and eventually consistent state replication. Third, it incorporates a content discovery layer that leverages distributed hash tables (DHTs) together with an optimized RPC protocol for efficient model synchronization. By integrating these components, Lattica delivers a complete protocol stack for sovereign, resilient, and scalable AI systems that operate independently of centralized intermediaries. It is directly applicable to edge intelligence, collaborative reinforcement learning, and other large-scale distributed machine learning scenarios .
\end{abstract}


\begin{figure*}[!ht]
    \centering
    \includegraphics[width=1.05\linewidth]{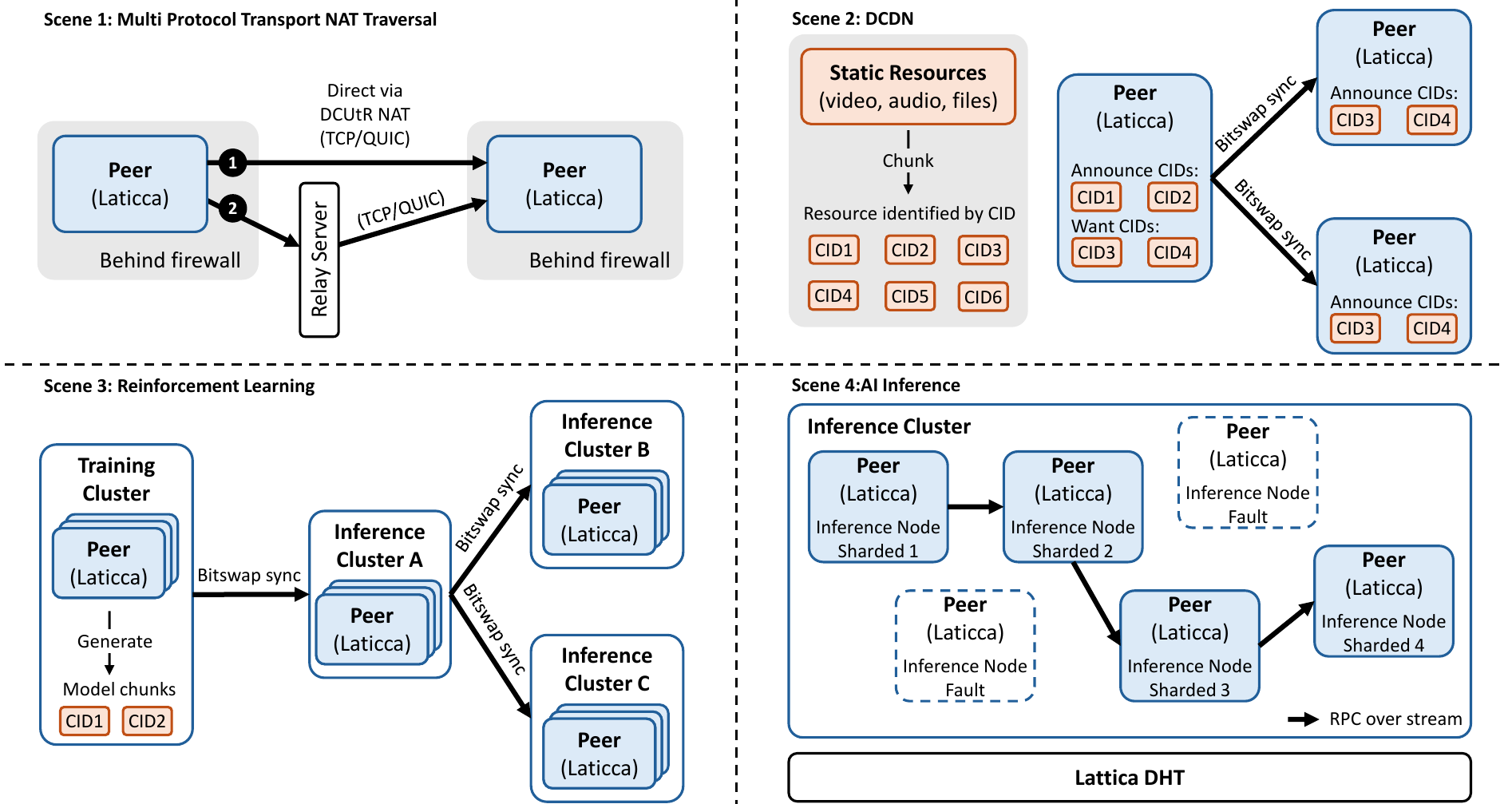}
\caption{\textbf{Lattica—Four Scenarios:} 
(1) Multi-protocol NAT traversal using Direct Connection Upgrade through Relay (DCUtR) over TCP and QUIC, with relay fallback; (2) Decentralized Content Delivery Network (CDN) in which static resources are chunked, content-identifier (CID) addressed, and synchronized via the Bitswap protocol; (3) Reinforcement learning pipeline where a training cluster generates model chunks (e.g., CID1, CID2) and synchronizes them with inference clusters A–C; (4) Sharded AI inference over the Lattica Distributed Hash Table (DHT) using Remote Procedure Call (RPC) streams with fault-tolerant shard nodes.}
    \label{fig:scenes}
\end{figure*}

\section{Introduction}

Modern machine learning workloads increasingly require distributed computing across many nodes, yet existing solutions reveal a tension between centralized orchestration and peer-to-peer decentralization~\cite{chen2022fairness}. For instance, Ray~\cite{moritz2018ray} is a widely adopted distributed computing framework that provides a general-purpose cluster runtime for ML tasks. While Ray effectively scales training and serving within data centers, it relies on centrally managed clusters with head nodes and reliable interconnects. In contrast, recent projects such as Hivemind~\cite{hivemind} demonstrate fully peer-to-peer deep learning on volunteer networks, leveraging a distributed hash table (DHT)~\cite{stoica2001chord}. Collectively, these developments underscore the need for communication infrastructure explicitly designed for decentralized AI—capable of supporting peer discovery, connectivity across Network Address Translators (NATs)~\cite{ford2005peer}, high-throughput data exchange, and fault tolerance in unreliable network environments.

Concurrently, the distributed systems community has developed foundational building blocks for decentralized communication~\cite{tanenbaum2017distributed}. A prominent example is \textbf{libp2p}, a modular peer-to-peer (P2P) networking stack originating from the InterPlanetary File System (IPFS) project~\cite{benet2014ipfs,libp2p}, which introduces efficient mechanisms for content addressing and distribution. In IPFS, cryptographic content identifiers (CIDs)~\cite{benet2014ipfs} are used to name data, while a distributed hash table (DHT) maps each CID to peers that can provide the corresponding content. Peers retrieve data via the Bitswap protocol, requesting blocks from any available neighbor~\cite{ipfsBitswap}. This design effectively enables a decentralized content delivery network (CDN)~\cite{pathan2008taxonomy}, where files or model parameters can be replicated and served by multiple sources without reliance on a central server. However, applying such content networks to AI workloads introduces new requirements: beyond file transfers~\cite{reddi2021vision}, training and inference demand fast, fine-grained communication among nodes, including parameter updates, gradient exchanges, and remote procedure calls for model serving.

Motivated by these gaps, we propose \textbf{Lattica}, a decentralized cross-NAT communication framework specifically designed to support distributed AI workloads. Unlike existing content networks, which primarily optimize for bulk file distribution~\cite{pathan2008taxonomy,nygren2010akamai,krishnamurthy2001design}, Lattica provides a generalized communication substrate capable of sustaining low-latency, high-throughput interactions required for modern AI training and inference. By addressing connectivity, consistency, data distribution, and execution challenges in a unified manner, Lattica lays the foundation for a new generation of resilient, sovereign, and scalable decentralized applications.  

The design of Lattica directly addresses the core obstacles of decentralized AI systems. For \textbf{peer-to-peer connectivity}, Lattica integrates advanced NAT traversal mechanisms that establish reliable communication channels even when peers are located behind firewalls or network address translators~\cite{ford2005peer,schollmeier2001definition}. To ensure \textbf{distributed data consistency}, it incorporates a decentralized store based on conflict-free replicated data types (CRDTs)~\cite{shapiro2011conflict,shapiro2012crdt}, which allow all nodes to converge on a verifiable and consistent state despite intermittent connectivity or untrusted participants. For the \textbf{distribution of large artifacts}~\cite{shoeybi2019megatron,dean2012large}, Lattica leverages content-addressed storage combined with distributed hash tables (DHTs) and the Bitswap protocol to propagate massive AI models efficiently and reliably across geographically dispersed peers. Finally, to enable \textbf{execution on constrained devices}, Lattica supports sharded inference and distributed training, partitioning workloads so that resource-limited nodes can participate collaboratively in computation~\cite{han2016deep}. The Lattica system is thus designed as an integrated solution to these diverse challenges, providing a foundational protocol that enables decentralized AI applications without requiring developers to repeatedly address the same set of infrastructural problems.  

Beyond its architectural contributions, Lattica opens new possibilities for practical deployment. For instance, it can support \textbf{edge intelligence}, where sensor-rich but resource-constrained devices collectively train and serve models without reliance on cloud providers. It also enables \textbf{collaborative reinforcement learning} across multiple organizations, where training clusters exchange model updates through decentralized coordination while preserving autonomy. Moreover, large-scale \textbf{federated or volunteer computing} initiatives can benefit from Lattica’s robust communication substrate to synchronize models across thousands of peers in unreliable networks. These scenarios illustrate how Lattica transforms decentralized AI from a conceptual possibility into an operational reality.

\section{System Architecture}
\begin{spacing}{0.98}
Lattica’s architecture is organized as a layered peer-to-peer (P2P) networking stack with bindings for high-level AI applications. At its foundation lies a Rust-based core built atop \textbf{libp2p}~\cite{libp2p}, which provides modular capabilities for decentralized networking. These include multi-transport support (TCP, QUIC, WebSocket, WebRTC)~\cite{benet2014ipfs,rfc9000}, secure encrypted channels~\cite{rfc9000}, peer identity~\cite{schollmeier2001definition}, and a plug-in framework for discovery and routing~\cite{maymounkov2002kademlia,stoica2001chord}. Above this core, Lattica integrates decentralized systems components—including peer discovery~\cite{castro2002secure}, DHT-based routing~\cite{maymounkov2002kademlia,stoica2001chord}, content addressing~\cite{benet2014ipfs}, pub-sub messaging~\cite{eugster2003many}, NAT traversal~\cite{ford2005peer}, and RPC~\cite{grpc}—that together form a comprehensive communication substrate. These capabilities are exposed to AI applications through language-specific SDKs (e.g., a Python SDK via Foreign Function Interface), allowing researchers to easily integrate Lattica into distributed training or inference pipelines. As shown in Figure~\ref{fig:scenes}, the architecture supports three representative usage scenarios: connectivity, content distribution, and AI inference.  

\textbf{Connectivity: Multi-Protocol NAT Traversal.} Establishing connections between peers behind NATs or firewalls is a critical challenge in decentralized networking. Lattica addresses this challenge with a multi-protocol NAT traversal mechanism orchestrated by a rendezvous service and leveraging \textbf{libp2p}’s NAT traversal modules. It supports transport over TCP and QUIC and dynamically negotiates fallback strategies such as relay servers and hole punching. If two peers cannot establish a direct connection (e.g., both behind NATs), Lattica employs libp2p’s AutoNAT service to discover each peer’s public reachability and, if necessary, engage a circuit relay as an intermediary. Once a connection is established, either directly or via relay, it is upgraded with authenticated encryption (Noise protocol~\cite{perrin2018noise} or TLS 1.3~\cite{rfc8446}, as provided by libp2p) to ensure confidentiality and integrity. By supporting multiple transports—TCP for broad compatibility, QUIC for low-latency multiplexing~\cite{rfc8825}, and WebRTC~\cite{webrtc2021w3c,loreto2014webrtc} for browser-based clients—Lattica can operate reliably across heterogeneous environments.  

\textbf{Content-Addressed Data Synchronization.} For data dissemination, Lattica adopts content-addressed storage and decentralized synchronization inspired by \textbf{IPFS}~\cite{ipfsBitswap}. Each data block is identified by a content identifier (CID)~\cite{multiformatsCID}, computed as a cryptographic hash of its contents. Peers announce and discover CIDs using a distributed hash table (DHT) based on the Kademlia algorithm~\cite{maymounkov2002kademlia}, which enables $O(\log N)$ lookup in a network of size $N$. Once a provider is located via the DHT (or via a rendezvous service for expedited discovery), data is retrieved through a BitSwap-like protocol. This mechanism allows, for example, a training node to publish model updates as sets of CID-identified blocks, which multiple worker or edge nodes can fetch concurrently from any peer storing them. In effect, this creates a decentralized CDN that reinforcement learning and inference clusters exploit to rapidly distribute new model versions.  

\textbf{RPC and Streaming for Training and Inference.} For interactive computation, Lattica offers a Protobuf-based RPC mechanism~\cite{protobuf,grpc} implemented over libp2p streams. It supports both request–response and streaming interactions. The request–response mode is designed for metadata and control-plane operations, such as health probes, shard placement, or model version queries, where low latency and idempotent retries are critical. The streaming mode is intended for tensors and other long-lived flows: multiplexed streams are established with adaptive backpressure~\cite{jacobson1988congestion,reactivestreams}, where writers monitor acknowledgments and queue depths while readers utilize zero-copy buffers to minimize CPU overhead. The SDK provides shard-aware client stubs that route requests across inference shards and transparently retry failed calls by resolving alternate providers through the DHT, thereby preserving availability. Built on a Rust core with Foreign Function Interfaces (FFI)~\cite{chaudhuri2005ffi,rustFFI} bindings (e.g., Python), the SDK abstracts networking complexity while exposing content and RPC APIs that align naturally with ML workflows and complement higher-level training and inference frameworks. This functionality is also highlighted in Figure~\ref{fig:scenes}, which illustrates how RPC streams enable sharded inference across distributed environments.
\end{spacing}

\section{Application Scenarios}
\begin{spacing}{0.95}
Beyond its architectural contributions, Lattica enables deployment across diverse real-world settings. The following scenarios illustrate its applicability to distributed AI systems operating outside the constraints of centralized cloud infrastructures.  

\textbf{Edge Intelligence.} In many Internet-of-Things (IoT) deployments, resource-constrained devices such as cameras, sensors, or mobile robots must execute learning and inference locally~\cite{shi2016edge}. For example, a smart-city deployment may rely on hundreds of roadside cameras to collaboratively train traffic flow prediction models. With Lattica, these devices can form a peer-to-peer mesh that disseminates updated models without a central server, ensuring robustness even in environments with intermittent connectivity.  

\textbf{Collaborative Reinforcement Learning.} Multiple organizations often wish to collaborate on reinforcement learning tasks without sacrificing autonomy or exposing private infrastructure. Consider logistics companies training warehouse robots: each organization can operate its own training cluster, but periodically exchange updated policies or value functions with others. Using Lattica’s DHT-based coordination and RPC streaming, these clusters can synchronize models through decentralized communication while avoiding dependence on a single orchestration point~\cite{zhang2021multiagent}.  

\textbf{Federated and Volunteer Computing.} Large-scale collaborative efforts, such as federated learning or volunteer-based model training, require efficient synchronization across thousands of geographically dispersed peers. A concrete example is a medical federated learning consortium where hospitals contribute model updates from sensitive datasets~\cite{kairouz2021federated}. With Lattica, these updates can be disseminated using content addressing and retrieved by other participants, even when nodes operate behind NATs or unstable networks. Similarly, volunteer computing projects such as SETI@home demonstrate the feasibility of harnessing global peers for large-scale computation~\cite{anderson2002seti}. Lattica’s peer-to-peer substrate enables such efforts to scale beyond the limitations of traditional client–server infrastructures.  

These scenarios highlight how Lattica transforms decentralized AI from a conceptual design into an operational reality, supporting robust, scalable, and collaborative learning in heterogeneous environments.
\end{spacing}

\section{Evaluation}
We evaluated Lattica’s core subsystems through preliminary experiments, focusing on NAT traversal success and RPC performance. In tests with peers deployed behind diverse NAT types~\cite{ford2005peer}, Lattica’s hole punching achieved direct peer-to-peer connectivity in roughly 70\% of attempts, while the remaining cases fell back to relay intermediaries. This success rate is comparable to prior measurements of \textbf{libp2p}’s NAT traversal capabilities~\cite{libp2p,ford2005peer}, indicating that Lattica can connect most nodes directly and still reach all nodes via relays when direct traversal fails. As a result, a robust global peer mesh can be maintained even in the presence of hard-to-penetrate firewalls.  

We also benchmarked Lattica’s Remote Procedure Call (RPC) throughput under various network conditions. Table~\ref{tab:throughput} summarizes the throughput (in queries per second, QPS) achieved for 1000 concurrent RPC calls with small (128\,B) and large (256\,KB) message payloads, using 4-core, 8\,GB machines on 10\,Gbps networks. In the best-case scenario, where client and server were colocated, Lattica sustained up to $\sim$10k QPS for 128\,B payloads. When nodes communicated across distant regions over the public Internet, throughput for 128\,B messages dropped to $\sim$1.2k QPS. For 256\,KB payloads, throughput reached about 850 QPS on a single host, versus about 110 QPS across continents. Intermediate scenarios (e.g., within the same region) achieved performance between these extremes. These results are consistent with prior observations on RPC performance in distributed systems~\cite{dean2012large,verma2015borg}, demonstrating that Lattica’s RPC mechanism can sustain high request rates in favorable conditions while maintaining usable performance across wide-area links, where bandwidth and latency constraints are most pronounced.

\begin{table}[!ht]
\centering
\caption{Lattica RPC throughput at 1000 concurrent calls (queries per second).}
\label{tab:throughput}
\begin{tabular}{lcc}
\toprule
\midrule
Network Scenario & 128\,B payload & 256\,KB payload \\
\midrule
Local (same host)        & 10000   & 850   \\
Same region (LAN)        & 8000    & 600   \\
Same region (WAN)        & 3000    & 280   \\
Inter-continent (WAN)    & 1200    & 110   \\
\midrule
\bottomrule
\end{tabular}
\end{table}


\section{User Study}
\begin{spacing}{0.9}
To complement the system-level benchmarks, we conducted a small-scale user study to evaluate Lattica’s usability and practical utility in real-world AI workflows. We recruited twelve participants, including graduate students and researchers with prior experience in distributed machine learning frameworks such as Ray~\cite{moritz2018ray} and PyTorch Distributed~\cite{paszke2019pytorch}. Each participant was tasked with deploying a distributed training or inference job using Lattica’s Python SDK on a cluster of heterogeneous machines spanning different NAT environments.

\textbf{Study Design.} The study consisted of two phases. In the \emph{deployment phase}, participants followed minimal documentation to install and configure Lattica, establish peer connectivity, and run a provided reinforcement learning pipeline across 6--10 nodes. In the \emph{evaluation phase}, participants adapted one of their own workloads (e.g., image classification, federated aggregation, or inference serving) to run on Lattica, reporting both technical challenges and performance observations.

\textbf{Results.} Participants successfully completed the deployment task in under 45 minutes on average, with 10 out of 12 reporting that NAT traversal and peer discovery were handled transparently without manual configuration. In their custom workloads, participants highlighted that Lattica’s content-addressed synchronization simplified sharing large model artifacts across nodes, while the RPC API allowed seamless integration with existing PyTorch training loops. Common feedback included requests for tighter integration with high-level ML frameworks and improved monitoring dashboards.

\textbf{Takeaways.} The study suggests that Lattica can be adopted by users with prior distributed ML experience with relatively low learning overhead. Its abstraction of connectivity and data synchronization reduces the engineering burden typically associated with cross-NAT deployments. At the same time, future iterations should improve usability features such as workload orchestration, logging, and visualization tools to better align with user expectations from established ML ecosystems.
\end{spacing}

\section{Conclusion}

Lattica integrates established peer-to-peer (P2P) primitives into a purpose-built substrate for distributed AI in adversarial and heterogeneous networks. Its multi-protocol NAT traversal, content-addressed storage, and dual-plane RPC framework address the challenges of transferring large artifacts while preserving the responsiveness required for tight control loops. Guided by the deployment and security considerations outlined in this work, practitioners can leverage Lattica to enable sharded inference and collaborative training across diverse environments with predictable performance and operational clarity.

\bibliography{neurips_2025}

\begin{thebibliography}{41}
\providecommand{\natexlab}[1]{#1}
\providecommand{\url}[1]{\texttt{#1}}
\expandafter\ifx\csname urlstyle\endcsname\relax
  \providecommand{\doi}[1]{doi: #1}\else
  \providecommand{\doi}{doi: \begingroup \urlstyle{rm}\Url}\fi

\bibitem[Chen et~al.(2022)Chen, Liao, Tian, Wang, and Yu]{chen2022fairness}
Zheyi Chen, Weixian Liao, Pu~Tian, Qianlong Wang, and Wei Yu.
\newblock A fairness-aware peer-to-peer decentralized learning framework with heterogeneous devices.
\newblock \emph{Future Internet}, 14\penalty0 (5):\penalty0 138, 2022.

\bibitem[Moritz et~al.(2018)Moritz, Nishihara, Wang, Tumanov, Liaw, Liang, Elibol, Yang, Paul, Jordan, et~al.]{moritz2018ray}
Philipp Moritz, Robert Nishihara, Stephanie Wang, Alexey Tumanov, Richard Liaw, Eric Liang, Melih Elibol, Zongheng Yang, William Paul, Michael~I Jordan, et~al.
\newblock Ray: A distributed framework for emerging $\{$AI$\}$ applications.
\newblock In \emph{13th USENIX symposium on operating systems design and implementation (OSDI 18)}, pages 561--577, 2018.

\bibitem[Ryabinin et~al.(2020)Ryabinin, Borzunov, Diskin, Gusev, Mazur, Plokhotnyuk, Bukhtiyarov, Samygin, Sinitsin, and Chumachenko]{hivemind}
Max Ryabinin, Alexander Borzunov, Michael Diskin, Anton Gusev, Denis Mazur, Vsevolod Plokhotnyuk, Alexey Bukhtiyarov, Pavel Samygin, Anton Sinitsin, and Artem Chumachenko.
\newblock {H}ivemind: {D}ecentralized {D}eep {L}earning in {P}y{T}orch, April 2020.
\newblock URL \url{https://github.com/learning-at-home/hivemind}.

\bibitem[Stoica et~al.(2001)Stoica, Morris, Karger, Kaashoek, and Balakrishnan]{stoica2001chord}
Ion Stoica, Robert Morris, David Karger, Frans Kaashoek, and Hari Balakrishnan.
\newblock Chord: A scalable peer-to-peer lookup service for internet applications.
\newblock In \emph{Proceedings of the 2001 Conference on Applications, Technologies, Architectures, and Protocols for Computer Communications (SIGCOMM)}, pages 149--160. ACM, 2001.

\bibitem[Ford et~al.(2005)Ford, Srisuresh, and Kegel]{ford2005peer}
Bryan Ford, Pyda Srisuresh, and Dan Kegel.
\newblock Peer-to-peer communication across network address translators.
\newblock \emph{USENIX Annual Technical Conference}, pages 179--192, 2005.

\bibitem[Tanenbaum and Van~Steen(2017)]{tanenbaum2017distributed}
Andrew~S Tanenbaum and Maarten Van~Steen.
\newblock \emph{Distributed systems}.
\newblock CreateSpace Independent Publishing Platform, 2017.

\bibitem[Benet(2014)]{benet2014ipfs}
Juan Benet.
\newblock Ipfs - content addressed, versioned, p2p file system.
\newblock \url{https://ipfs.io/ipfs/QmR7GSQM93Cx5eAg6a6jx5dhFAj2fjwHaJgN7iePZmqCio}, 2014.
\newblock White paper.

\bibitem[{libp2p Project}(2019)]{libp2p}
{libp2p Project}.
\newblock libp2p: A modular peer-to-peer networking stack.
\newblock \url{https://libp2p.io/}, 2019.
\newblock Accessed: 2025-09-29.

\bibitem[{IPFS Project}(2017)]{ipfsBitswap}
{IPFS Project}.
\newblock Bitswap protocol specification.
\newblock \url{https://github.com/ipfs/specs/tree/main/bitswap}, 2017.
\newblock Specification, accessed 2025-09-29.

\bibitem[Pathan et~al.(2008)Pathan, Buyya, and Vakali]{pathan2008taxonomy}
Al-Mukaddim~Khan Pathan, Rajkumar Buyya, and Athena Vakali.
\newblock A taxonomy and survey of content delivery networks.
\newblock In \emph{Content Delivery Networks}, pages 33--77. Springer, 2008.
\newblock \doi{10.1007/978-3-540-77887-5_2}.

\bibitem[Reddi et~al.(2021)Reddi, Cheng, Kanter, Mattson, Schmuelling, and Wu]{reddi2021vision}
Vijay~Janapa Reddi, Christine Cheng, David Kanter, Peter Mattson, Guenther Schmuelling, and Carole-Jean Wu.
\newblock The vision behind mlperf: Understanding ai inference performance.
\newblock \emph{IEEE Micro}, 41\penalty0 (3):\penalty0 10--18, 2021.

\bibitem[Nygren et~al.(2010)Nygren, Sitaraman, and Sun]{nygren2010akamai}
Erik Nygren, Ramesh~K. Sitaraman, and Jennifer Sun.
\newblock The akamai network: A platform for high-performance internet applications.
\newblock In \emph{ACM SIGOPS Operating Systems Review}, volume~44, pages 2--19. ACM, 2010.
\newblock \doi{10.1145/1842733.1842736}.

\bibitem[Krishnamurthy et~al.(2001)Krishnamurthy, Wills, and Zhang]{krishnamurthy2001design}
Balachander Krishnamurthy, Craig~E. Wills, and Yinglian Zhang.
\newblock On the design and performance of internet content distribution networks.
\newblock \emph{IEEE Internet Computing}, 5\penalty0 (4):\penalty0 68--80, 2001.
\newblock \doi{10.1109/4236.935182}.

\bibitem[Schollmeier(2001)]{schollmeier2001definition}
R{\"u}diger Schollmeier.
\newblock A definition of peer-to-peer networking for the classification of peer-to-peer architectures and applications.
\newblock In \emph{Proceedings First International Conference on Peer-to-Peer Computing}, pages 101--102. IEEE, 2001.
\newblock \doi{10.1109/P2P.2001.990434}.

\bibitem[Shapiro et~al.(2011)Shapiro, Pregui{\c{c}}a, Baquero, and Zawirski]{shapiro2011conflict}
Marc Shapiro, Nuno Pregui{\c{c}}a, Carlos Baquero, and Marek Zawirski.
\newblock Conflict-free replicated data types.
\newblock In \emph{Stabilization, Safety, and Security of Distributed Systems (SSS)}, pages 386--400. Springer, 2011.
\newblock \doi{10.1007/978-3-642-24550-3_29}.

\bibitem[Shapiro et~al.(2012)Shapiro, Pregui{\c{c}}a, Baquero, and Zawirski]{shapiro2012crdt}
Marc Shapiro, Nuno Pregui{\c{c}}a, Carlos Baquero, and Marek Zawirski.
\newblock Conflict-free replicated data types.
\newblock \emph{ACM Transactions on Computer Systems (TOCS)}, 28\penalty0 (1):\penalty0 1--38, 2012.
\newblock \doi{10.1145/2136349.2136350}.

\bibitem[Shoeybi et~al.(2019)Shoeybi, Patwary, Puri, LeGresley, Casper, and Catanzaro]{shoeybi2019megatron}
Mohammad Shoeybi, Mostofa Patwary, Raul Puri, Patrick LeGresley, Jared Casper, and Bryan Catanzaro.
\newblock Megatron-lm: Training multi-billion parameter language models using model parallelism.
\newblock In \emph{Proceedings of the International Conference on Machine Learning (ICML)}, pages 4550--4560. PMLR, 2019.

\bibitem[Dean et~al.(2012)Dean, Corrado, Monga, Chen, Devin, Le, Mao, Ranzato, Senior, Tucker, Yang, and Ng]{dean2012large}
Jeffrey Dean, Greg Corrado, Rajat Monga, Kai Chen, Matthieu Devin, Quoc~V. Le, Mark~Z. Mao, Marc'Aurelio Ranzato, Andrew Senior, Paul Tucker, Ke~Yang, and Andrew~Y. Ng.
\newblock Large scale distributed deep networks.
\newblock In \emph{Advances in Neural Information Processing Systems (NeurIPS)}, volume~25, 2012.

\bibitem[Han et~al.(2016)Han, Mao, and Dally]{han2016deep}
Song Han, Huizi Mao, and William~J. Dally.
\newblock Deep compression: Compressing deep neural networks with pruning, trained quantization and huffman coding.
\newblock \emph{International Conference on Learning Representations (ICLR)}, 2016.

\bibitem[Iyengar and Thomson(2021)]{rfc9000}
Jana Iyengar and Martin Thomson.
\newblock Quic: A udp-based multiplexed and secure transport.
\newblock RFC 9000, Internet Engineering Task Force (IETF), 2021.
\newblock URL \url{https://www.rfc-editor.org/rfc/rfc9000}.
\newblock Accessed: 2025-09-29.

\bibitem[Maymounkov and Mazieres(2002)]{maymounkov2002kademlia}
Petar Maymounkov and David Mazieres.
\newblock Kademlia: A peer-to-peer information system based on the xor metric.
\newblock In \emph{Proceedings of the 1st International Workshop on Peer-to-Peer Systems (IPTPS)}, pages 53--65. Springer, 2002.

\bibitem[Castro et~al.(2002)Castro, Druschel, Ganesh, Rowstron, and Wallach]{castro2002secure}
Miguel Castro, Peter Druschel, Atul Ganesh, Antony Rowstron, and Dan~S. Wallach.
\newblock Secure routing for structured peer-to-peer overlay networks.
\newblock In \emph{Proceedings of the ACM Symposium on Operating Systems Principles (SOSP)}, pages 299--314. ACM, 2002.
\newblock \doi{10.1145/1060289.1060315}.

\bibitem[Eugster et~al.(2003)Eugster, Felber, Guerraoui, and Kermarrec]{eugster2003many}
Patrick~Th. Eugster, Pascal~A. Felber, Rachid Guerraoui, and Anne-Marie Kermarrec.
\newblock The many faces of publish/subscribe.
\newblock \emph{ACM Computing Surveys}, 35\penalty0 (2):\penalty0 114--131, 2003.
\newblock \doi{10.1145/857076.857078}.

\bibitem[{Google}(2015)]{grpc}
{Google}.
\newblock grpc: A high performance, open source universal rpc framework.
\newblock \url{https://grpc.io/}, 2015.
\newblock Accessed: 2025-09-29.

\bibitem[Perrin(2018)]{perrin2018noise}
Trevor Perrin.
\newblock The noise protocol framework.
\newblock \url{https://noiseprotocol.org/noise.pdf}, 2018.
\newblock Revision 34, accessed 2025-09-29.

\bibitem[Rescorla(2018)]{rfc8446}
Eric Rescorla.
\newblock The transport layer security (tls) protocol version 1.3.
\newblock RFC 8446, Internet Engineering Task Force (IETF), 2018.
\newblock URL \url{https://www.rfc-editor.org/rfc/rfc8446}.
\newblock Accessed: 2025-09-29.

\bibitem[Alvestrand(2021)]{rfc8825}
Harald Alvestrand.
\newblock Overview: Real-time protocols for browser-based applications (webrtc).
\newblock RFC 8825, Internet Engineering Task Force (IETF), 2021.
\newblock URL \url{https://www.rfc-editor.org/rfc/rfc8825}.
\newblock Accessed: 2025-09-29.

\bibitem[{W3C}(2021)]{webrtc2021w3c}
{W3C}.
\newblock Webrtc 1.0: Real-time communication between browsers.
\newblock \url{https://www.w3.org/TR/webrtc/}, 2021.
\newblock Candidate Recommendation, accessed 2025-09-29.

\bibitem[Loreto and Romano(2014)]{loreto2014webrtc}
Salvatore Loreto and Simon~Pietro Romano.
\newblock Real-time communications in the web: Issues, achievements, and ongoing standardization efforts.
\newblock \emph{IEEE Communications Surveys \& Tutorials}, 16\penalty0 (4):\penalty0 1856--1870, 2014.
\newblock \doi{10.1109/COMST.2014.2320139}.

\bibitem[{Multiformats Project}(2017)]{multiformatsCID}
{Multiformats Project}.
\newblock Content identifiers (cid) specification.
\newblock \url{https://github.com/multiformats/cid}, 2017.
\newblock Specification, accessed 2025-09-29.

\bibitem[{Google}(2008)]{protobuf}
{Google}.
\newblock Protocol buffers: Google's data interchange format.
\newblock \url{https://developers.google.com/protocol-buffers}, 2008.
\newblock Accessed: 2025-09-29.

\bibitem[Jacobson(1988)]{jacobson1988congestion}
Van Jacobson.
\newblock Congestion avoidance and control.
\newblock In \emph{Proceedings of the ACM SIGCOMM Conference}, pages 314--329. ACM, 1988.
\newblock \doi{10.1145/52324.52356}.

\bibitem[{Reactive Streams Project}(2015)]{reactivestreams}
{Reactive Streams Project}.
\newblock Reactive streams: An initiative to provide a standard for asynchronous stream processing with non-blocking back pressure.
\newblock \url{https://www.reactive-streams.org/}, 2015.
\newblock Accessed: 2025-09-29.

\bibitem[Chaudhuri and Foster(2005)]{chaudhuri2005ffi}
Avik Chaudhuri and Jeffrey~S. Foster.
\newblock Foreign function interfaces for functional languages.
\newblock In \emph{Proceedings of the 10th ACM SIGPLAN International Conference on Functional Programming (ICFP)}, pages 163--175. ACM, 2005.
\newblock \doi{10.1145/1086365.1086390}.

\bibitem[B.(2015)]{rustFFI}
Sean B.
\newblock The rust ffi omnibus.
\newblock \url{http://jakegoulding.com/rust-ffi-omnibus/}, 2015.
\newblock Accessed: 2025-09-29.

\bibitem[Shi et~al.(2016)Shi, Cao, Zhang, Li, and Xu]{shi2016edge}
Weisong Shi, Jie Cao, Quan Zhang, Youhuizi Li, and Lanyu Xu.
\newblock Edge computing: Vision and challenges.
\newblock \emph{IEEE Internet of Things Journal}, 3\penalty0 (5):\penalty0 637--646, 2016.
\newblock \doi{10.1109/JIOT.2016.2579198}.

\bibitem[Zhang et~al.(2021)Zhang, Yang, and Basar]{zhang2021multiagent}
Kaiqing Zhang, Zhuoran Yang, and Tamer Basar.
\newblock Multi-agent reinforcement learning: A selective overview of theories and algorithms.
\newblock \emph{Handbook of Reinforcement Learning and Control}, pages 321--384, 2021.
\newblock \doi{10.1007/978-3-030-60990-0_11}.

\bibitem[Kairouz et~al.(2021)Kairouz, McMahan, Avent, Bellet, Bennis, Bhagoji, Bonawitz, Charles, Cormode, Cummings, et~al.]{kairouz2021federated}
Peter Kairouz, H.~Brendan McMahan, Brendan Avent, Aur{\'e}lien Bellet, Mehdi Bennis, Arjun~Nitin Bhagoji, Keith Bonawitz, Zachary Charles, Graham Cormode, Rachel Cummings, et~al.
\newblock Advances and open problems in federated learning.
\newblock \emph{Foundations and Trends in Machine Learning}, 14\penalty0 (1--2):\penalty0 1--210, 2021.
\newblock \doi{10.1561/2200000083}.

\bibitem[Anderson et~al.(2002)Anderson, Cobb, Korpela, Lebofsky, and Werthimer]{anderson2002seti}
David~P. Anderson, Jeff Cobb, Eric Korpela, Matt Lebofsky, and Dan Werthimer.
\newblock Seti@home: An experiment in public-resource computing.
\newblock \emph{Communications of the ACM}, 45\penalty0 (11):\penalty0 56--61, 2002.
\newblock \doi{10.1145/581571.581573}.

\bibitem[Verma et~al.(2015)Verma, Pedrosa, Korupolu, Oppenheimer, Tune, and Wilkes]{verma2015borg}
Abhishek Verma, Luis Pedrosa, Madhukar Korupolu, David Oppenheimer, Eric Tune, and John Wilkes.
\newblock Large-scale cluster management at google with borg.
\newblock In \emph{Proceedings of the European Conference on Computer Systems (EuroSys)}, pages 1--17. ACM, 2015.
\newblock \doi{10.1145/2741948.2741964}.

\bibitem[Paszke et~al.(2019)Paszke, Gross, Massa, Lerer, Bradbury, Chanan, Killeen, Lin, Gimelshein, Antiga, et~al.]{paszke2019pytorch}
Adam Paszke, Sam Gross, Francisco Massa, Adam Lerer, James Bradbury, Gregory Chanan, Trevor Killeen, Zeming Lin, Natalia Gimelshein, Luca Antiga, et~al.
\newblock Pytorch: An imperative style, high-performance deep learning library.
\newblock In \emph{Advances in Neural Information Processing Systems (NeurIPS)}, volume~32, pages 8024--8035, 2019.

\end{thebibliography}
\bibliographystyle{unsrtnat}
\end{document}